# Sex differences in intimate relationships

Vasyl Palchykov[1], Kimmo Kaski[1], Janos Kertész[1,2], Albert-László Barabási[2,3] & Robin I.M. Dunbar[4]

[1]Department of Biomedical Engineering and Computational Science (BECS), Aalto University School of Science, P.O. Box 12200, FI-00076, Finland, [2]Institute of Physics BME, Budapest, Budafoki ut 8., H-1111, Center for Complex Networks Research, Northeastern University, Boston, MA 02115, [4]Institute of Cognitive & Evolutionary Anthropology, University of Oxford, 64 Banbury Road, Oxford OX2 6PN, UK

**Social networks based on dyadic relationships are fundamentally important for understanding of human sociality. However, we have little understanding of the dynamics of close relationships and how these change over time. Evolutionary theory suggests that, even in monogamous mating systems, the pattern of investment in close relationships should vary across the lifespan when post-weaning investment plays an important role in maximising fitness. Mobile phone data sets provide a unique window into the structure and dynamics of relationships. We here use data from a large mobile phone dataset to demonstrate striking sex differences in the gender-bias of preferred relationships that reflect the way the reproductive investment strategies of both sexes change across the lifespan, i.e. women's shifting patterns of investment in reproduction and parental care. These results suggest that human social strategies may have more complex dynamics than previously assumed and a life-history perspective is crucial for understanding them.**

Social relationships, and in particular pairbonds, are the outcome of individuals' decisions about whom to invest their available social capital in. Such decisions typically reflect a choice between the payoffs offered by alternative candidates. However, in monogamous species, and especially those that live in multi-generational families, investment strategies may vary across the lifespan as a function of the individual's changing reproductive circumstances – notably the impact of constraints such as the risk of death or the cessation of active reproduction[1,2]. In species like humans, where menopause truncates female reproductive activity and investment in offspring typically continues into adulthood, evolutionary theory would predict that investment in relationships should vary across the lifetime as a function of the trade off between the relative opportunities for personal reproduction versus (grand-)parental investment. In this respect, evolutionary theory would also predict significant contrasts in the social strategies of the two sexes as a function of the differences in their reproductive strategies. However, studying human social relationships in any detail on a large scale has proved unusually difficult. The bottom-up approach adopted by social psychologists and sociologists has commonly been limited by sample size, while the more recent top-down social network analysis approach inevitably suffers from a lack of detail about the individuals involved[3,4]. More importantly, most large-scale network studies have tended to treat relationships as static, and ignore the fact that social relationships are dynamic and change over time, at the very least on the scale of a lifetime.

In humans, homophily (a tendency for individuals who share traits to preferentially form relationships) has emerged as an important organizing principle of social behaviour[5,6]. In most such cases, studies of homophily have focused on psychological or social traits such as personality, interests, hobbies, and religious or political views. However, there is evidence that homophily may also arise through a tendency for close friendships to be gender-biased[7,8] and we exploit this to explore the changing patterns of relationship investment across the lifespan. We use a cross-sectional analysis of a very large mobile phone database to investigate gender preferences in close friendships: i) to test the hypothesis that preferences in the choice of the "best friend" are gender-biased (homophilic with respect to gender); and ii) to investigate how these preferences change over the lifespan. We focus our attention on the three most preferred friends, as indexed by the frequency of contact. Several studies have demonstrated that frequency of contact is a reliable index of emotional closeness in relationships[9,10], and these datasets confirm that frequency of contact by telephone and other digital media (text, email) correlates significantly with frequency of face-to-face contact ($p \ll 0.0001$ in each case, N=1006 and N=8967, respectively). Recent research also reveals that personal social networks are hierarchically structured[11,12], having a layer-like structure with distinct differences in emotional closeness and frequency of contact with alters in the different layers, with an inner core of ~ 5 alters who between them account for about half our total social time[9,13].

**Results**

For our study we used the large-scale hashed mobile phone dataset from a single mobile service provider in a specific European country[14–16]. The dataset covers a seven-month period and includes 1.95 billion calls and 489 million text messages. Carrying out initial data filtering we arrive at $N \approx 3.2$ million subscribers, of whom about 1.8 million are males and about 1.4 million females. Finally, we performed some additional data filtering to remove obviously erroneous records in the dataset as described in the Methods section.

We define the "best friend" of a given subscriber $i$ as the alter that $i$ is most frequently in contact with, counting both the number of calls and text messages; the "second best friend" is then the alter that $i$ is in contact with next most frequently, and the "third best friend" is the next most frequently contacted individual, etc. Restricting ourselves to pairs of subscribers for both of whom we have age and gender information gives us 1.19 million ego/best-friend pairs, 0.80 million ego/second-best-friend pairs and 0.66 million ego/third-best-friend pairs. These numbers of ego/friend pairs are smaller than what would come out if the total number of subscribers in the sample could be used. The reduction is related to the described restrictions with the dataset and may introduce uncorrelated randomness, which, however, we do not expect to have any significant influence on the main conclusions of this study.

For our analysis, we identify the gender of each subscriber $i = 1...N$ by the variable $g_i$, such that $g_i = 1$ for males and $g_i = -1$ for females. We define the average gender as $\langle g \rangle = \frac{1}{N}\sum_i g_i$, where the summation is taken over all subscribers. Since $\langle g \rangle \approx 0.13$ for the whole dataset, there is an imbalance in favour of male alters. The average gender $\langle f \rangle$ of the ego's "best friend" is defined as $\langle f \rangle = \frac{1}{N_f} \sum_i f_i$, where $f_i$ stands for the gender of the "best friend" of subscriber $i$ and in the summation $i$ runs over $N_f$ pairs of subscribers with known gender and age information. With the overall $\langle f \rangle \approx -0.01$, the "best friends" are almost perfectly balanced. Compared that to $\langle g \rangle \approx 0.13$ indicates that there is a strong bias in the selection due to some kind of gender correlation. (The balance of the egos and the best friends is depicted as a function of the age of the egos in Supplementary Fig. S1 online). Considering male and female subscribers separately, we reveal that the "best friends" are usually characterized by opposite genders with the average gender of the "best friend" having an overall value $\langle f \rangle = -0.26$ for males and $\langle f \rangle \approx 0.29$ for females.

In order to determine what this gender correlation is, we examine the average gender of the "best friend" as a function of ego's age, for male and female egos separately (Fig. 1a). It is apparent that until the age of about 50 years, both male and female egos prefer their "best friend" to be of the opposite gender, although this effect is strongest for 32 year old males and 27 year old females, yielding peak values of $\langle f \rangle \approx -0.41$ and $\langle f \rangle \approx 0.46$ for males and females, respectively.

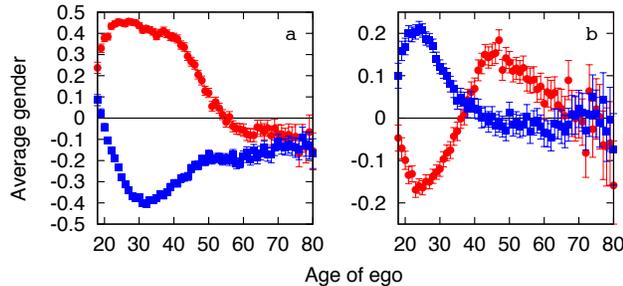

**Figure 1. Gender correlations between best friends.** A: Average gender of the "best friend" of an ego of specified age and gender (n male, female). B: Average gender of the "second best friend" of an ego of specified age and gender (n male, female). Error bars show confidence interval with significance level $\alpha = 0.05$. Note that these results are overall independent of the definition of the "best friend" (see Supplementary Fig. S3 online).

Notice that not only does the preference for an opposite-sex "best friend" kick in noticeably earlier for females than for males (~18 years vs. ~ 22, respectively), but females maintain a higher plateau value for much longer. Males exhibit a distinct and quite short-lived peak (of about 7 years as indicated by a 20% decline from the peak value); in contrast, women have a long, relatively higher male-biased plateau (of about 14 years, as defined by the onset of a 20% decline from the peak), after which the male "best friend" seems to be moved to the second place (Fig. 1b) and is replaced as "best friend" by a new (typically female) alter. While males' "best friends" remain slightly female-biased throughout their lives, women's only become so during their early 50s. The two sexes eventually converge on a slightly female-biased pattern at around 70 years of age.

The pattern for the "second best friend" (Fig. 1b) is a partial mirror image of the pattern for the "best friend". "Second best friends" are typically same-sex, reaching a sharp peak during the subjects' early 20s before falling away gradually to reverse the gender bias for individuals in their late 30s. The transition is sharper for women than for men: males exhibit a shallower decline, and settle at an asymptotic value very close to gender equality, whereas women show a striking reversal to a strong male-biased peak in their late 40s (and a steady decline back towards equality by the late 60s). We ran a similar analysis for the gender of the third, fourth, and fifth best friend and since the patterns are virtually identical to that in Fig. 1b (albeit with a strong male-bias in older age for both sexes), we present the results only in the Supplementary Fig. S4 online. The similarity of the plots for the second, third, fourth, and fifth best friends reinforce the contrast with the case for the "best friend", suggesting a more privileged status for the "best friend".

In Fig. 2 we illustrate these findings in the form of a network with links representing gender correlations between the "best friends". Red circles correspond to female, blue circles to male subscribers, and grey circles correspond to subscribers with unavailable age and gender information (or subscribers of other service providers). The thicknesses of the links (and the number) stand for the frequencies of contact, thus illustrating the emotional closeness between the pair of individuals. In addition, we have used the circle sizes to reflect the subscriber ages: the bigger the circle, the older the subscriber. Beside gender correlations, this local weighted network shows the age correlations between the "best friends". It can be seen that young people prefer the "best friend" to be of opposite gender and of the same age group. One can also see very distinct patterns in older individuals' communication patterns, namely that a 50 year old female subscriber has a young female (possibly a daughter) as the "best friend" and as the "second best friend" a male of her own age group (possibly her husband). What we also see in this case is that the "third best friend" is typically also of younger generation but male (possibly son). Note the very strong opposite-gender focus of relationships among 20 and 30 year olds, suggesting strong pairbond focus.

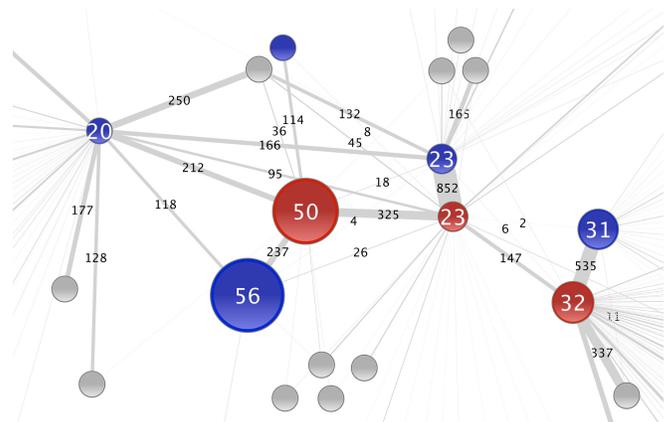

**Figure 2. A sample of local network between best friends.** A part of the network with a gender and age correlations. Blue circles correspond to male and red circles to female subscribers. Circle sizes reflect subscriber ages: the bigger the circle, the older the subscriber. Grey circles correspond to subscribers, whose gender and age information is not available in our data set.

We can see these effects more clearly by considering age correlations between "best friends". In Fig. 3, we show age distributions of the "best friends" for both male and female egos aged 25 and 50 years. (Similar plots for the intervening age cohorts are given in Supplementary Figs. S5-S8 online.) On this finer scale analysis, some additional patterns emerge. The distributions for friends of both genders turn out to be bimodal, with one maximum at around ego's own age and the other at an age difference of approximately 25 years, i.e. one generation apart. The maxima at ego's own age are opposite-gender biased, and most likely identify a male partner for female egos and vice versa. The maxima at the 25 year age difference (i.e. the generation gap) show a more balanced gender ratio, most likely identifying children and parents, respectively, for 50-and 25-year-old egos. In supplementary figures, the progression of this split can be seen very clearly in the profiles for the intervening age cohorts.

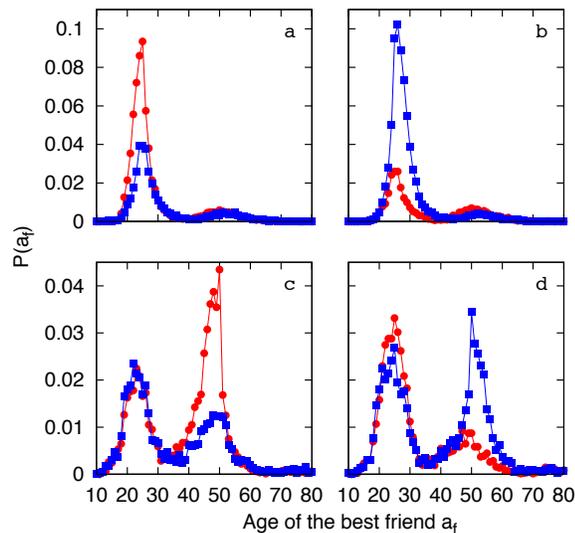

**Figure 3. Distributions of the "best friends" by age.** The distributions of the "best friends" by age for 25 years old (a) male and (b) female egos. In c and d we show similar distributions for 50 years old male and female egos, respectively. Red circles correspond to female "best friends" while blue squares to male "best friends". Each data point displays the probability that the "best friend" is of specified age and gender.

**Discussion**

On the assumption that mobile phone communication represents the most of important relationships of subscribers and that the strength of communication reflects the level of emotional closeness, these results allow us to draw four conclusions. First, women are more focused on opposite-sex relationships than men are during the reproductively active period of their lives, suggesting that they invest more heavily in creating and maintaining pairbonds than men do[17]. Second, as they age, women's attention shifts from their spouse to younger females, whom we assume, on the basis of the age difference, to be their daughters. This transition is relatively smooth and slow for women (perhaps taking about 15 years to reach its new asymptote at around age 60), and may reflect the gradual arrival of grandchildren. Third, women switch individuals around in their preference rankings much more than men do, suggesting that their relationships are more focused while men's are more diffuse. Men tend to keep a steadier pattern over a longer period, maintaining a preference for placing their spouse in pole position across time and a striking tendency to maintain a very even gender balance in the second position. If the latter represent offspring, then the data suggest a strong lack of discrimination. In contrast, women tend to switch individuals from one position to another in a more exaggerated way, perhaps reflecting shifts in their allegiances as their reproductive strategies switch more explicitly from mate choice to personal reproduction to (grand-)parental investment, particularly after age 40. Women's gender-biases thus tend to be stronger than men's, seemingly because their patterns of social contact are strongly driven by the changes in the patterns of reproductive investment across the lifespan. Women's stronger inclination toward parental and grandparental investment is attested to by the striking contrast with the pattern exhibited by men: men's gender-biases for both best and second/third best friends show much less evidence for any preference for contacting children. Indeed, the younger (25-year-old) peak for 50-year-old men is half that for women and shows a more even sex balance, whereas that for women is strongly biased in favour of female alters (presumably, daughters) (Fig. 3c,d), presumably reflecting the maternal grandmother investment effects previously noted in demographic studies[18,19]. Finally, fourth, our results provide strong evidence for the importance of female matrilineal relationships in human social organisation. There has been a tendency to emphasise the importance of male-male relationships in an essentially patrilineal form of social organisation as defining human sociality[20], but our results tend to support the claim that mother-daughter relationships play a particularly seminal role in structuring human social relationships irrespective of dispersal pattern, as has been suggested by some sociological studies[21].

While, inevitably, our analyses pool together large numbers of individuals, and so lose some of the richness of the original data, nonetheless we have been able to demonstrate striking patterns in mobile phone usage data that reflect shifts in relationship preferences as a function of the way the reproductive strategies of the two sexes change across the lifespan. Such patterns have not been noted previously, and our findings suggest that there are novel opportunities for exploiting large network datasets of this kind if the right kinds of questions are asked. Aside from this purely methodological aspect, our analyses identify striking sex differences in the social and reproductive strategies of the two sexes that have not been previously identified. We have also been able to demonstrate a marked sex difference in investment in relationships during the period of pairbond formation, suggesting that women invest much more heavily in pairbonds than do men. Though previously suspected[17], this suggestion has proved particularly difficult to test. Finally, we should note that our analyses have focused on the simple presence/absence of contacts: further insights into sex and age related differences in human communication patterns might be gained by analysing both the link directionalities or asymmetries in who initiates communication and the structural and temporal motifs of such directional networks.

**Methods**

Although the majority of the recorded subscribers are between 20-60 years of age, the size of our sample and the fact that it is a saturation sample means that even those age classes that are relatively rare (e.g. between 60 and 80 years of age) are still represented by a sizeable sample (~5000 individuals) (see Supplementary Fig. S1).

**Initial data filtering**

In order to filter out spurious effects like accidental or wrong number events as well as professional (e.g. call centre) calls, we considered calls and text messages only between individuals that had at least one reciprocated contact (i.e. a contact in each direction). Among the 6.8 million subscribers of this provider who meet this criterion, we consider only those whose gender and age are both known, and for whom only a single subscription is registered. Finally, we assume that subscriber and real user of mobile phone are identical if no obvious deviations are observed. Below, we describe such example observed in the dataset and the way of its filtration.

**Additional data filtering**

Taking into account the distribution of birthdays over a year, we can assume that the probability of an ego having a best friend of his or her own age and that of having a best friend who differs from his/her age by one-year should not be significantly different. In fact, a high peak for ego's own precise age and gender is observed in the initially filtered dataset. The existence of this peak contradicts the above assumption and may be caused by multiple subscriptions registered for a single person but not recorded in the database: this might happen if multiple phones are registered to one individual, but are used by children or partners, during the course of which they call ego. Hence, this artifact needs to be eliminated. Different approaches to the data filtering may be applied. Our approach consists of predicting the number $n$ of real ego/friend pairs of equal age $a$ and gender $g$. In the current study, we suppose that this number $n$ is equal to the bigger of the numbers represented by ego/friend pairs where friends are one year older or younger than the ego considered. In order to obtain reliable results for the average gender of the best friend only $n$ randomly chosen pairs of ego/friend of equal age $a$ and gender $g$ are considered. Finally, this procedure is repeated for each age and gender group of an ego. The same approach is applied to the pairs of ego and second best friend of equal age and gender. The effect of this additional filtering is demonstrated in supplementary figure S2.

**Error estimation**

Finite sample sizes cause errors as depicted in Fig. 1 by error bars, as a result of error estimation. For convenience we here concentrate on the best friend only by considering egos of specified age $a$ and gender $g$. Among remaining $n$ pairs of ego/friend, $m$ friends are males and $n-m$ friends are females. Since gender variable has only two possible values, the distribution of the best friend's gender is bimodal. To estimate the errors, we introduce a quantity $x$ that gives a fraction of males among the best friends and perform Bayesian inference [22]. Conditional distribution function $p(x|m,n)$ for the fraction of males $x$ given $m$ appearances of males among $n$ friends is proportional to $p(x|m,n) \propto x^m (1-x)^{n-m}$ and reaches its maximum at $x^* = m/n$. For a given significance level $\alpha$ one may estimate credible interval for variable $x \in (x_{\min}, x_{\max})$, where $x_{\min}$ and $x_{\max}$ are defined in the way that the probabilities for $x$ to be larger or smaller than $x^*$ are equal to $(1-\alpha)/2$. Writing this condition in terms of incomplete regularized beta function $I_z(p,q) = B_z(p,q)/B_1(p,q)$, where $B_z(p,q) = \int_0^z x^{p-1}(1-x)^{q-1} dx$ we arrive at the following equation for $x_{\min}$:

$$I_{x_{\min}}(m+1, n-m+1) = I_{m/n}(m+1, n-m+1) - (1-\alpha)/2. \qquad (1)$$

The value of $x_{\min}$ is obtained from equation (1) as inverse incomplete beta function $I_z^{-1}(p,q)$. We obtain the value of $x_{\max}$ in a similar way. The credible interval for the fraction of male best friends, $x$, determines the credible interval for average gender as $\langle f \rangle \in (2x_{\min}-1, 2x_{\max}-1)$. These calculations are repeated for egos of each age and gender group.

**Acknowledgements**

Financial support from EU's 7th Framework Program's FET-Open to ICTeCollective project no. 238597 and by the Academy of Finland, the Finnish Center of Excellence program, project no. 129670, and TEKES (FiDiPro) are gratefully acknowledged.

**Author contributions**

All authors were involved in planning, designing and realizing the research. VP devised and realized statistical analysis of the empirical data and prepared graphical presentations in the manuscript. All authors contributed to manuscript preparation in all of its stages.

**Additional information**

**Supplementary information** accompanies this paper at http://www.nature.com/scientificreports

**Competing financial interests:** The authors declare no competing financial interests.

# Supplementary information

## Sex differences in intimate relationships

by Vasyl Palchykov, Kimmo Kaski, Janos Kertész, Albert-László Barabási & Robin I.M. Dunbar

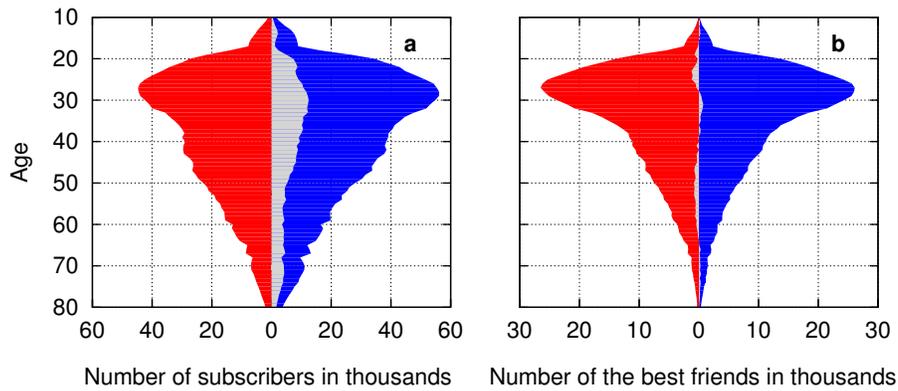

**Supplementary Figure S1. Population pyramids.** Panel **a**: distribution of subscribers by age. Blue and red lines correspond to males and females, respectively. The number of males exceeds the number of females for each age group. The difference between these numbers is shown by grey colour and gives rise to overall average gender $\langle g \rangle = 0.13$. Panel **b**: distribution of the egos' best friends by age. Blue and red lines correspond to male and female best friends while grey lines give the difference between them. The set of the best friends is very slightly female-biased with average gender $\langle f \rangle = -0.01$.

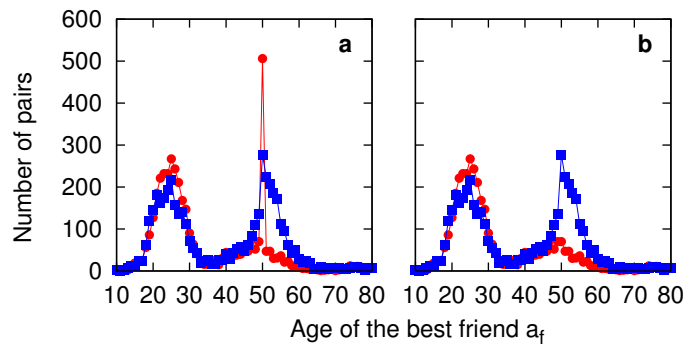

**Supplementary Figure S2. Additional data filtering.** The distributions of the best friend's age $a_f$ for 50 years old female egos. Panel **a** shows the distribution for original dataset. The distribution shows high peak at ego's own age and gender, which may be caused by multiple subscriptions registered for a single person if the real users of these different phones are his/her family members. Panel **b** shows the same distributions after filtering.

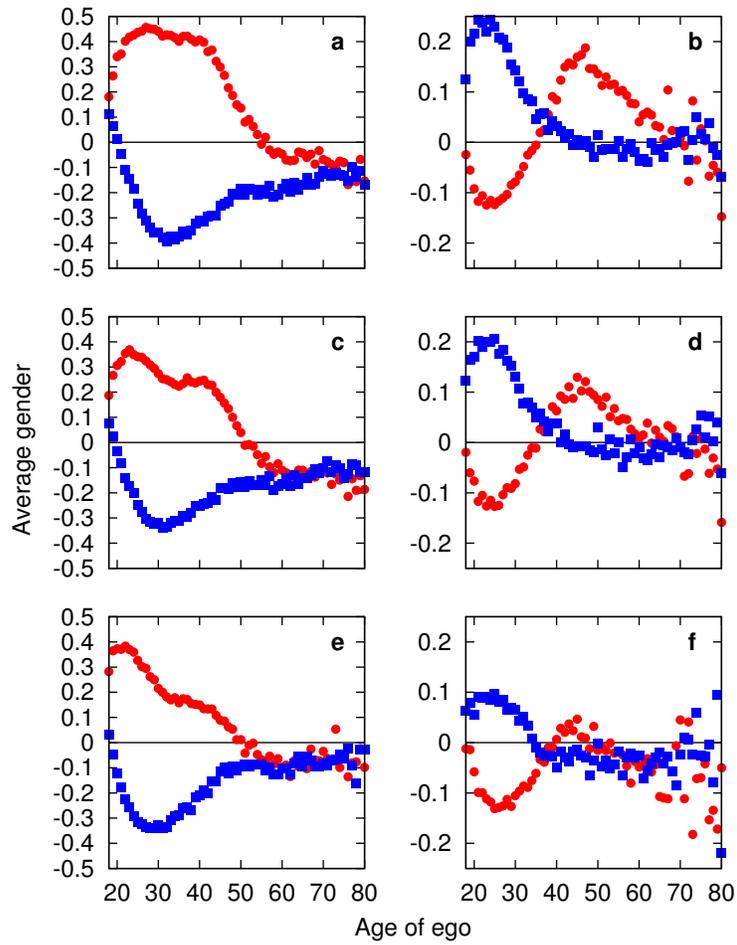

**Supplementary Figure S3.** Average gender of the best friend (left column) and the second best friend (right column) as a function of ego's age. Different rows correspond to different basis of the best friend definition: panels **a**, **b** (only number of calls); **c**, **d** (total call duration); **e**, **f** (number of text messages) between subscribers. All these results show generic features for the gender of corresponding friend.

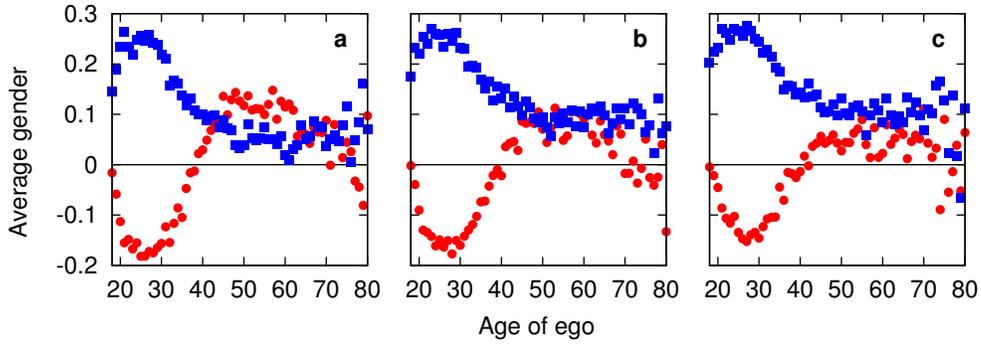

**Supplementary Figure S4.** Average gender of the ego's third best friend (panel **a**), fourth best friend (panel **b**), and fifth best friend (panel **c**) of specified age and gender. Blue squares correspond to males, red circles to females.

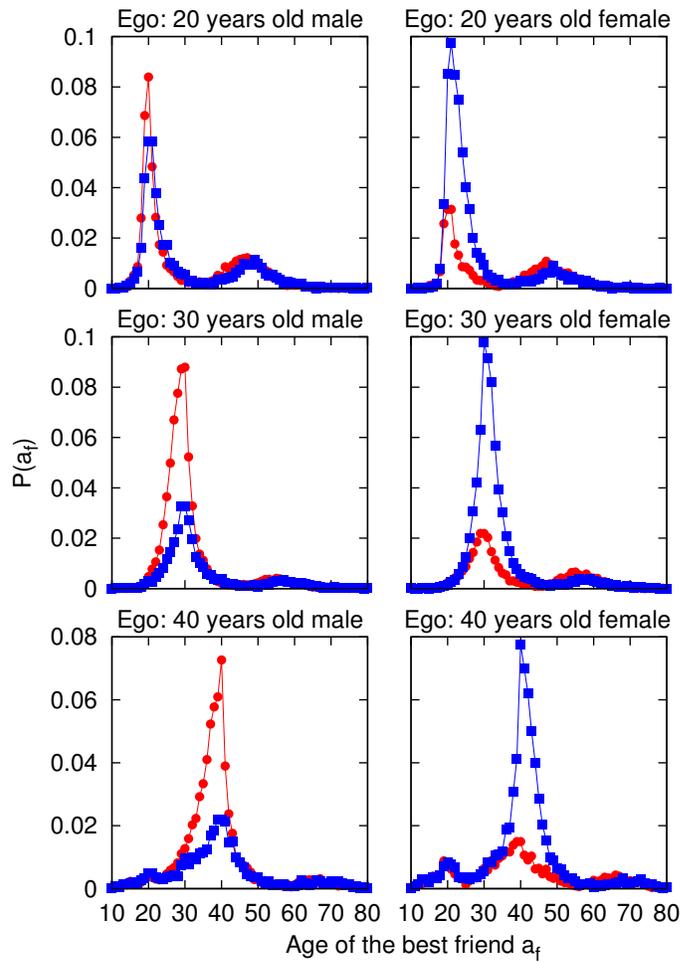

**Supplementary Figure S5.** The distributions of the best friends by age for male and female egos of fixed age of 20, 30, and 40 years. In each figure, two (or even three) maxima as well as asymmetry in opposite gender ages around ego's age are evident. Each data point displays the probability that the best friend is of specified age and gender.

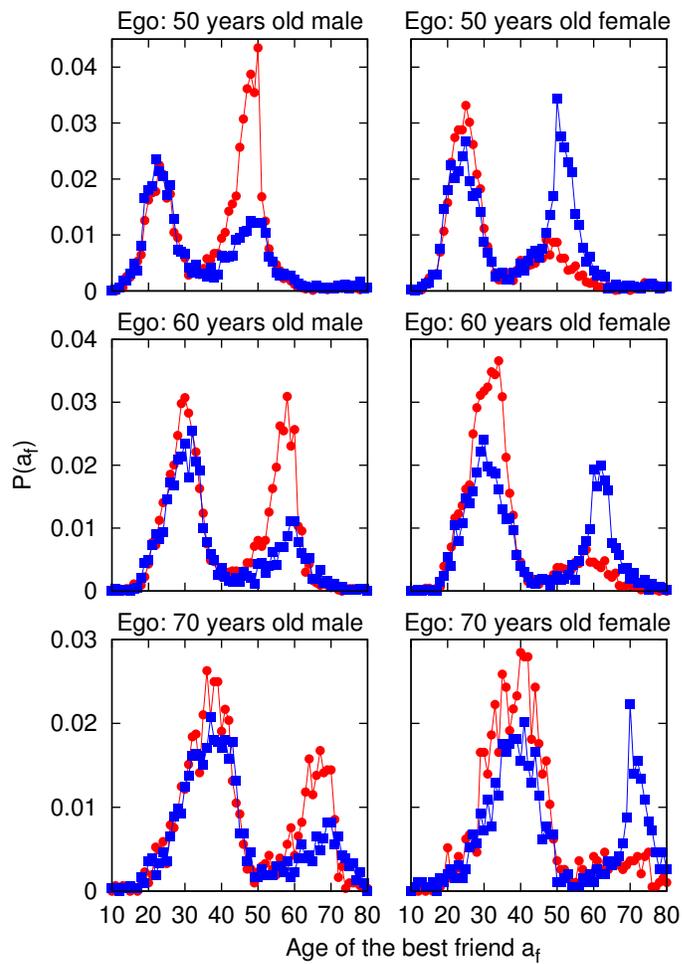

**Supplementary Figure S6.** The distributions of the best friends by age for male and female egos of fixed age of 50, 60, and 70 years. The increasing amount of communication with younger generation is observed. Each data point displays the probability that the best friend is of specified age and gender.

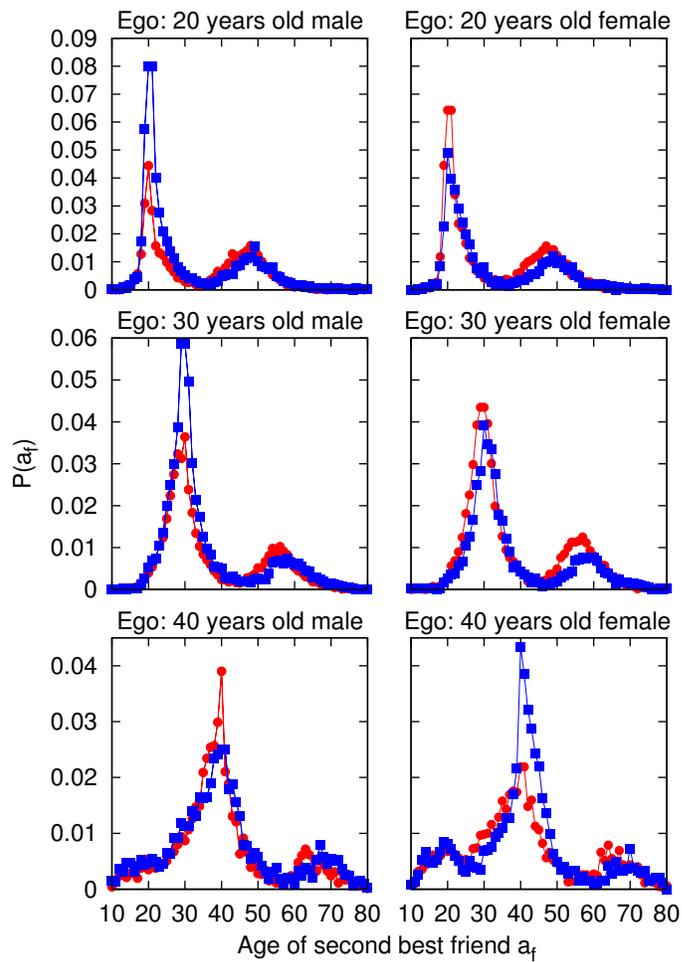

**Supplementary Figure S7.** The distribution of the second best friends by age for male and female egos of different age cohorts. The profiles are quite similar to those for best friends, but with striking differences in the gender of the friend: for younger egos, the second best friend of similar age to ego is more typically of the same gender as ego, whereas for older egos the second best friend of the same age as ego is more often of the opposite gender.

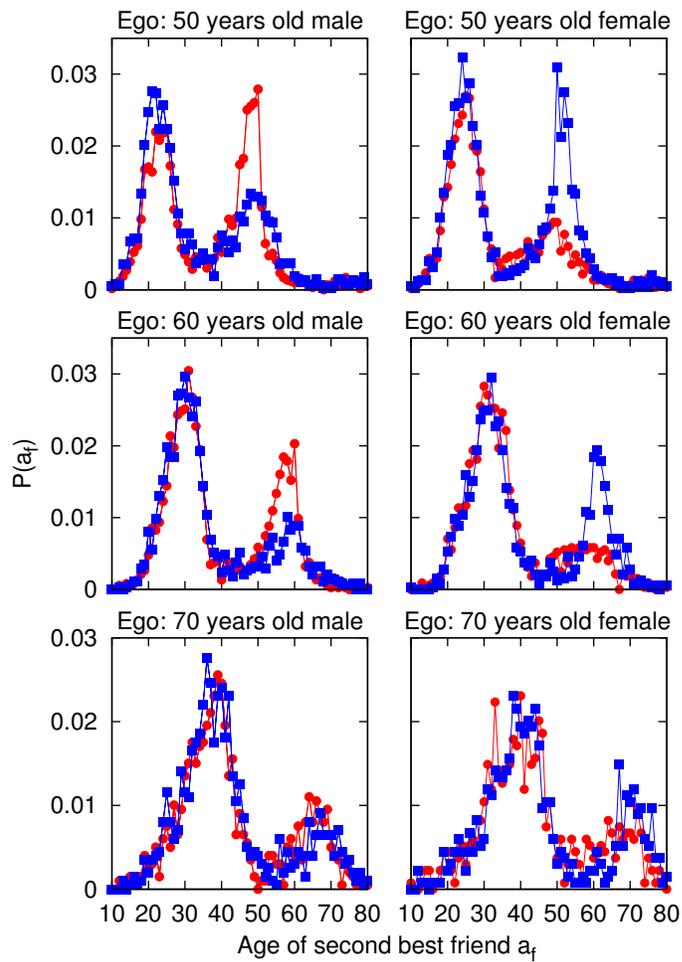

**Supplementary Figure S8.** The distribution of the second best friends by age for male and female egos of different age cohorts. Each data point displays the probability that the second best friend is of specified age and gender.